\documentclass[aps,10pt,prd,twocolumn,preprintnumbers,amsmath,amssymb,nofootinbib,superscriptaddress,a4paper,showpacs]{revtex4-1}

\usepackage[breaklinks]{hyperref}
\usepackage[english]{babel}
\usepackage[utf8]{inputenc}
\usepackage{graphicx}
\usepackage{color}
\usepackage{amsfonts,amsthm}
\usepackage{bm,bbm}

\def\lp{\ell_{\rm Pl}}

\newcommand{\ee}{\end{equation}}
\newcommand{\ba}{\begin{eqnarray}}
\newcommand{\ea}{\end{eqnarray}}
\def\bs{\begin{subequations}}
\def\es{\end{subequations}}

\def\com{\color{magenta}}
\def\cob{\color{blue}}
\newcommand{\book}[5]{\emph{#1} (#2, #3, #4, #5)}
\newcommand{\oarX}[1]{\href{http://arxiv.org/abs/#1}{{\ttfamily\com arXiv:#1}}}
\newcommand{\arX}[1]{\href{http://arxiv.org/abs/#1}{{\ttfamily\com arXiv:#1}}}
\newcommand{\doin}[6]{\href{http://dx.doi.org/#1}{{\cob #2 #3 {\bf #4}, #5 (#6)}}}
\newcommand{\doinn}[5]{\href{http://dx.doi.org/#1}{{\cob #2 {\bf #3}, #4 (#5)}}}
\newcommand{\doij}[5]{\href{http://dx.doi.org/#1}{{\cob #2 #3 (#5) #4}}}

\newcommand{\procsinm}[5]{in \emph{#1}, ed.\ by #2 (#3, #4, #5)}

\newcommand{\tia}[1]{}

\def\lp{\ell_{\rm Pl}}

\newcounter{listcounter}

\begin{document}

\title{Quantum coordinate operators: why space-time lattice is fuzzy}

\author{Suddhasattwa Brahma}
\email{suddhasattwa.brahma@gmail.com}
\affiliation{Center for Field Theory and Particle Physics, Fudan University, 200433 Shanghai, China}
\affiliation{Asia Pacific Center for Theoretical Physics, Pohang 37673, Korea}

\author{Antonino Marcian\`o}
\email{marciano@fudan.edu.cn}
\affiliation{Center for Field Theory and Particle Physics, Fudan University, 200433 Shanghai, China}

\author{Michele Ronco}
\email{michele.ronco@roma1.infn.it}
\affiliation{Dipartimento di Fisica, Universit\`a di Roma ``La Sapienza", P.le A. Moro 2, 00185 Roma, Italy, EU}
\affiliation{INFN, Sez.~Roma1, P.le A. Moro 2, 00185 Roma, Italy, EU}

\begin{abstract}
\noindent
Working within the framework of Loop Quantum Gravity (LQG), we construct a set of three operators suitable for identifying coordinate-like quantities on a spin-network configuration. In doing so, we  rely on known properties of operators for angles, which are already well-known in the LQG literature. These operators are defined on the kinematical Hilbert space, in a background-independent fashion. Computing their action on coherent states, we are able to study some relevant properties such us the spectra, which are discrete. In particular, we focus on the algebra generated by quantum coordinates and, remarkably, it turns out that they do not commute. Interestingly, this may provide additional hints on how space-time noncommutativity could be realized in the context of LQG. The semiclassical regime, necessary to make contact with coordinates on manifolds, is also explored and, specifically, is given by the large-spin limit in which commutativity can be restored. Finally, building on well-established results, we discuss how it is possible to have regularization. 
\end{abstract}

\maketitle

\section{introduction}
\noindent Both the logical analysis of the quantum gravity problem \cite{dfr,padma,gara,alu,ngdam,amelino,Hooft,ven,kon,ellis}, as well as some technical results within formal approaches to handle it \cite{Ori09,Fousp,Smo17,Zwi09,rov07,thie1,Per13,GiSi,AGJL4,Dow13,LaR5,NiR,RSnax,ADKLW,BIMM,Tom97,Mod1, MarMod,BGKM,CaMo2,gacLRR,tH93,Car09,fra1,frafuzz,frafuzz1,frabh,NewRev,Car17}, suggest that, at a fundamental level, spacetime should have a `fuzzy', quantum nature, which is rather different from the smooth continuum manifolds we are used to at the classical level. Among the studied forms of spacetime quantization, place of pride is held by the hypothesis of having a noncommutative nature of the coordinates on the manifold \cite{dfr,dfr1,majgac,witten,ncfra}. Spacetime noncommutativity turned out to be useful in the characterization of spacetime `fuzziness', which one expects to become important to describe physics in higher curvature regimes. Indeed, different forms of spacetime noncommutativity can be rigorously derived from both string theory \cite{witten,szabo,gaume,freidST1,freidST2} and $3$D quantum gravity \cite{3dCosm,freidelivine,kowRos}. On the other hand, it still remains unclear if there is room for noncommutativity of spacetime coordinates, in the context of quantization of the local gauge group of symmetries, in one of the most studied quantum-gravity approaches, namely loop quantum gravity (LQG) \cite{rov07,thie1, ash1, fb, gio, AshBojoLew}. Moreover, within the LQG framework, an element of complexity for the analysis of nocommutativity is represented by the fact that, by construction, there are no manifold coordinates due to the intrinsic background independence of the theory. Nonetheless, few encouraging steps have been taken in this direction over the last decade thanks to the study of effective models \cite{3dCosm,freidelivine,kowRos,orititlas,bal1,bal2,girelli,girelli1,bianco,NClimLQG,polymerNC}. For instance, in Ref.~\cite{freidelivine} it was shown that the effective dynamics of matter fields coupled to $3$D quantum gravity reduces, after integration over the gravitational degrees of freedom, to a braided noncommutative quantum field theory enjoying a deformed Poincar\'e group of symmetries. More recently, in Ref.~\cite{NClimLQG}, it has been shown that a specific Planck-scale deformation of the Poincar\'e algebra, obtained in the zero-curvature regime of the Dirac algebra of constraints with holonomy corrections from LQG \cite{holo1,holo2,holo3}, is dual to the so-called $\kappa$-Minkowski noncommutative spacetime \cite{majgac,lukrue,majrue}, whose coordinates close Lie algebra-type commutators. Further support to such a result has been provided in Ref.~\cite{polymerNC} by studying the relativistic symmetries of polymer quantum mechanics \cite{poly1,poly2}, which have many aspects in common with symmetry-reduced LQG models \cite{bojoLRR}. In both cases, the analysis of the symmetries is as crucial as the description of the fuzziness of the quantum geometry in terms of the noncommutativity of the space-time coordinates. Also along this direction, there has been much development in the literature, both at the level of the characterization of the deformed symmetries in terms of their associated conserved quantities \cite{Agostini:2006nc, Arzano:2007gr,AmelinoCamelia:2007uy,AmelinoCamelia:2007wk,AmelinoCamelia:2007rn,Marciano:2008tva,Marciano:2010jm}, focusing on their phenomenological applications to astrophysics and cosmology \cite{Marciano:2010gq, AmelinoCamelia:2010zf,AmelinoCamelia:2012it,BRAM,UVdimLQG}, and at the level of the consequences for the Fock space of quantum field theories enjoying deformed symmetries \cite{Arzano:2007ef, AmelinoCamelia:2007zzb}, and the related deformation of the multi-particle states statistics \cite{Btheta1,Btheta2,Addazi:2017bbg}.

The main implication of our analysis is that a form of noncommutativity arises as a feature of the fuzziness of spin-network nodes at mesoscales. In other words, noncommutativity emerges in our work from coarse-graining at larger scales the microscopic texture of the geometry, which at the Planckian scale is, in turn, described non-perturbatively by quantum operators and states on a Hilbert space. Specifically, in the LQG approach we deploy here, information about the quantum geometry is encoded in the quantum numbers assigned to the states that form a basis in the kinematical Hilbert space. This is the basis of spin-network states supported on graphs $\Gamma$, which in turn are composed of $N$ nodes and $L$ links. The coarse-graining procedure we adopt amounts to grouping in three sets all the links emanating from a node of $\Gamma$, while the semiclassical limit, achieved by sending to infinity irreducible representations of $SU(2)$ assigned to links, necessitates the choice of the coherent spin-network states. Then, the emergence of noncommutativity is an effective phenomenon for the quantum geometry that is retrieved while evaluating the semi-classical limit of operators defined in LQG. Such a result may suggest that departures from the smooth commutativity of the space-time manifold coordinates  (and thus corresponding modifications to the Poincar\'e symmetries) could play a role in an intermediate regime of scales that is close enough to the Planckian regime.

The kinematical Hilbert space of LQG is constructed from abstract spin-network graphs embedded in a manifold. A variety of well-defined geometric operators can be recovered on this Hilbert space which, in turn, can be used to calculate the spectra of areas and volumes. In this way, one regains information regarding the geometry of the spatial manifold from these abstract spin-network states. On the other hand, a gap still exists in the literature on how to regain information regarding the position of some coordinate chart on this manifold from the kinematical Hilbert space. This would require defining some suitable operators on the Hilbert space,  that could serve as a sort of coordinate defined by relying only on spin-network data. In any case, in the appropriate classical limit, we would expect to recover the usual classical coordinates on the manifold. Of course, the quantum property of these operators can be significantly more exotic than their classical analogs, as we shall see in this paper.

More concretely, we introduce an operator that should provide a sort of notion of coordinates on the kinematical Hilbert space of LQG. In fact, as stated above, there is a well-known detailed analysis of the properties of geometric quantities such as areas \cite{rovSm,geomop1}, volumes \cite{rovSm,geomop1,ash2} and also lengths \cite{len1,len2,len3}, but very little is known about what happens to spacetime points or, to put it more precisely, if there exists an analogous procedure to also characterize coordinates. A rather renowned result in the LQG literature states that areas, volumes and lengths, when realized as well-defined quantum operators on the kinematical Hilbert space, indeed have the remarkable feature of possessing discrete spectra \cite{rovSm,rovSm1,bianca}. We here propose a straightforward way of defining an operator that mimics the three spatial coordinates. In order to define them, we use previous results \cite{major1,major2} defining angle operators on spin-network states, where the angle is identified by the two links converging at the same node of the network.  By using this operator for the direction of links, we are able to find a suitable definition for what we call, from now on, the  ``coordinate operator" (CO). The idea that noncommutativity might arise in LQG as a consequence of direction quantization was proposed in Ref.~\cite{roveu} few years ago. To some extent our work can be regarded as a concrete realization of that proposal. Using the action on semi-classical coherent states \cite{CSetera,CSbianchi,CSalesci}, we compute the algebra generated by COs and find out that they do not commute and, in particular, they close an algebra which vaguely represents the one generated by angular-momentum (but with important differences as illustrated later), as suggested in some of the first toy models for spacetime quantization appeared in the literature \cite{madore}. Spin-network coordinates close a sort of Lie algebroid generalization of the $SU(2)$ Lie algebra that resembles the so-called fuzzy sphere \cite{snyder,madore}, where the structure functions depend on half of the phase-space variables (namely the fluxes). Consistently, in the large spin $j$ limit, we recover classical commutative spatial coordinates. Coordinates defined only in terms of quantum directions require the introduction of an arbitrary parameter with dimensions of length which, classically, can be identified as the distance of the point we consider, from the origin. A second proposal is also introduced, and briefly discussed, in the Appendix A, recovering results that are qualitative similar to the ones deepened here. Our COs shall not be diffeomorphism-invariant since the very notion of coordinates on (even a classical) manifold depends on the choice of the chart. However, defining operators that are not diffeomorphism-invariant is common in LQG, see e.g. the case of the `length'-operator. As a matter of fact, hitherto the full diffeomorphism-invariance was not even recovered (because of the current failure of imposing the scalar constraint in a general set-up of pure gravity in LQG) within the case of the more common area and volume operators, which nevertheless are widely treated in the literature \cite{rovSm,rovSm1}. Consequently, since we shall be defining our COs on the kinematical Hilbert space, these operators cannot be `observables' in the nomenclature of Dirac. Nonetheless, the reason for developing the proposal of geometrical noncommutative quantities, including the angle operators as much as the coordinate operators we are about to introduce here, lies in the possibility of gaining intuition about the emergent deformation of symmetries, which in stead retains a physical and (experimentally) observable meaning \cite{Addazi:2017bbg}.\\

The classical phase space of canonical gravity is given in terms of the couple of conjugate variables $(A^i_a(x),E^{ai}(x))$ \cite{ash1}. The Ashtekar-Barbero variables $A^i_a(x)$ are $SU(2)-$valued connections embedded in a 3-manifold $\Sigma$, and are conjugate to the triads $E^{ai}(x)$ of density weight one. Their Poisson algebra is given by
\begin{equation}
\begin{split}
\{ A^i_a(x), A^j_b(y) \} = \{ E^{a}_i(x), E^{b}_j(y) \} = 0 \, , \\ \quad \{ A^i_a(x), E^b_j(y) \} = 8\pi\gamma \delta^{(3)}(x,y)\delta^b_a\delta^i_j \, .
\end{split}
\end{equation}
Here, according to the usual notation, $\gamma$ stands for the Barbero-Immirzi parameter \cite{fb,gio}. Quantizing canonical gravity in Ashtekar's formulation, it is a pre-requirement to turn to holonomies and fluxes as fundamental phase-space variables. This is due to the fact that, in general, not the connection directly, but rather its parallel transport, can be represented as a well-defined operator in LQG \cite{AshBojoLew}. Connections are replaced by their holonomies
\begin{equation}
h_e[A] = \mathcal{P}\exp \int_e dt \, \dot{e}_a A^i_a \tau_i \, ,
\end{equation}
in which $\tau_i = -i\sigma_i/2 $, $\sigma_i$ denoting the Pauli matrices, and $\dot{e}_a=d{e}_a/dt$, tangent to the curve ${e}_a(t)$. Through holonomies, a group element of $SU(2)$ is then associated to each edge $e$. At the same time, a densitized triad $E^a_i$ is smeared over a two surface $S$
\begin{equation}\label{flux}
F_i[S] = \int_S E^a_i \epsilon_{abc} dx^b \wedge dx^c \, ,
\end{equation}
in order to get flux variables. The introduction of fluxes and holonomies as, respectively, two and one dimensional integrals, is a crucial point in order to have a background independent formulation in which the spacetime metric does not enter explicitly. Holonomies and fluxes can be quantized consistently and, thus, every geometric operators can then be expressed in terms of these fundamental variables of the theory. They constitute the basic ingredients of the loop quantization procedure \cite{rovSm,ashLewetal}. The algebra of operators between a holonomy $h_e[A] $ and a flux $F_i(S)$ is given by
\begin{equation}
\label{fluxact}
\{ F_i(S), h_e[A] \} = 8\pi \gamma \tau_i h_e[A] \circ (e,S) \, ,
\end{equation}
where $\circ (e,S)$ is equal to $0$ if either the path $e$ does not intersect $S$ or if it lies entirely on $S$, while it is $\pm 1$ if there is a single intersection where its sign depends on the mutual orientation of $e$ and $S$.

Taking $N$ copies of the $SU(2)$ group one can construct cylindrical functions
\begin{equation}
\Psi_\Gamma [A] = \psi(h_{e_1}[A],...,h_{e_N}[A])
\end{equation}
over the graph $\Gamma$ composed by these $N$ edges. These functionals of holonomies are dense in the (kinematical) Hilbert space equipped with the Ashtekar-Lewandowski measure \cite{ashLew}. An orthonormal basis of this space is provided by spin-network states \cite{rov1,thie1}
\begin{equation}
\Psi_{\Gamma,j,i} [A] = \bigotimes_{n \subset \Gamma}v_{i_n} \bigotimes_{e \subset \Gamma} \mathcal{D}^{(j_e)}( h_e[A])  \, ,
\end{equation}
where $\Gamma$ is a closed graph embedded in the 3-manifold $\Sigma$, and its edges $e$ are labelled by irreducible representations $j_e$ of $SU(2)$. The edges converge in nodes $n$, to which one associates intertwiners $i_n$. These latter are taken in the tensor product of the representations of the edges that intersect at a given node. Each holonomy $h_e[A]$  is written in the associated representation  $j_e$ of the group given by $ \mathcal{D}^{(j_e)}( h_e[A])$. Holonomies act as multiplicative operators on spin-network states, while fluxes amount to derivatives.
\begin{figure}[h!]
\centering
\includegraphics[width=3in]{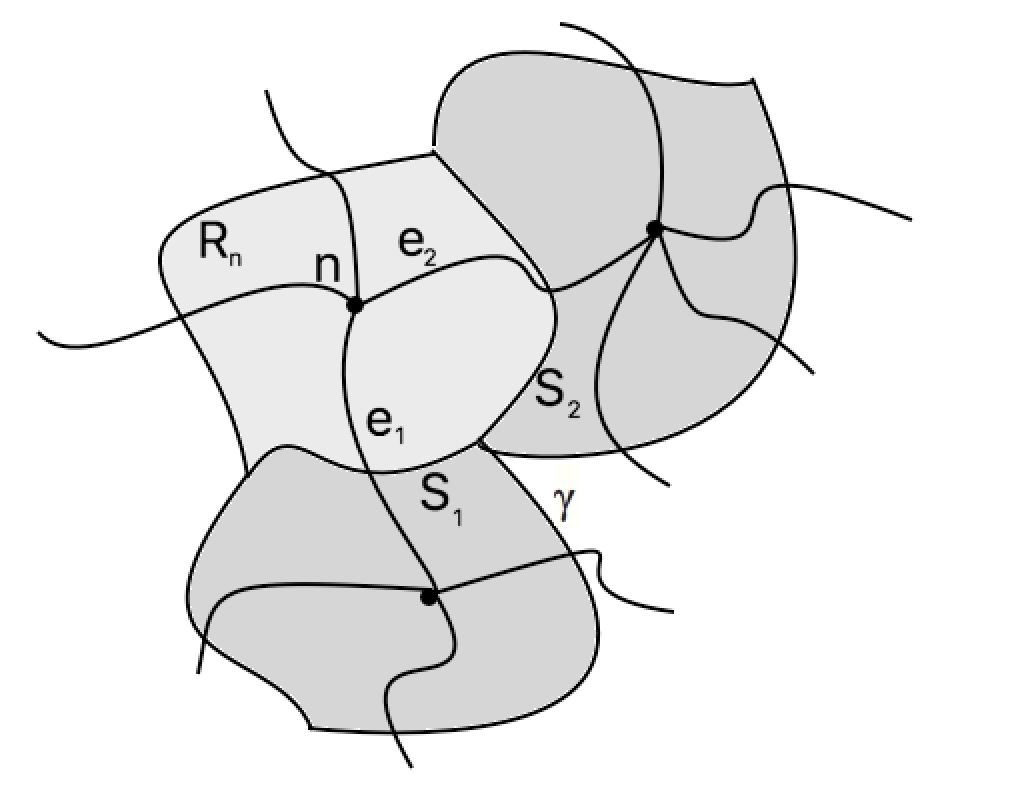}
\caption{The figure graphically represents a given spin network $\Gamma$ made up of three nodes. The region, named $R_n$, is the portion of space that is dual to the node $n$. Two of the four links converging at $n$, those labeled by $e_1$ and $e_2$, are then dual to the surfaces $S_1$ and $S_2$ respectively. The intersection of these $2d$ surfaces with the boundary of $R_n$ identifies the curve $\gamma$.  }
\label{sn}
\end{figure}
Thanks to the self-adjointness of the kinematical Hilbert space with the Ashtekar-Lewandowski measure and to the Peter-Weyl theorem providing the spin network basis, the theory can be quantized by giving an equivalent operatorial representation of Eq.  \eqref{fluxact}.


\section{Angle operators in LQG}

\begin{figure}[h!]
\centering
\includegraphics[width=3in]{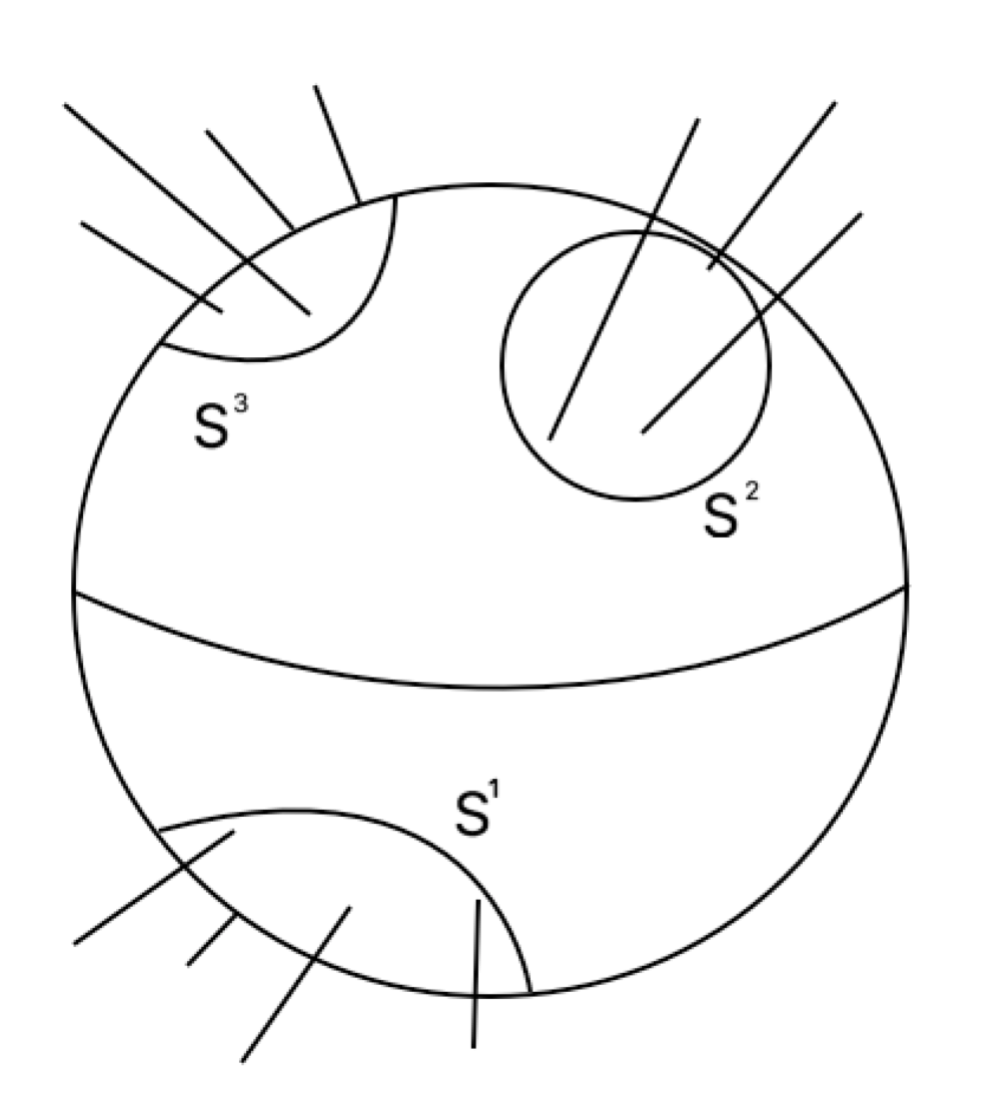}
\caption{The figure shows the way we group the links converging at a given node. We pick out three sets of links and gather them in three different ``total'' links, which we call $S_1$ and $S_2$ and $S_3$. Given such a construction, it is possible to define an operator for the angle between $S_1$ and $S_2$,  for the angle between $S_2$ and $S_3$, and for that between $S_1$ and $S_3$. }
\label{ang}
\end{figure}

\begin{figure}[h!]
\centering
\includegraphics[width=3in]{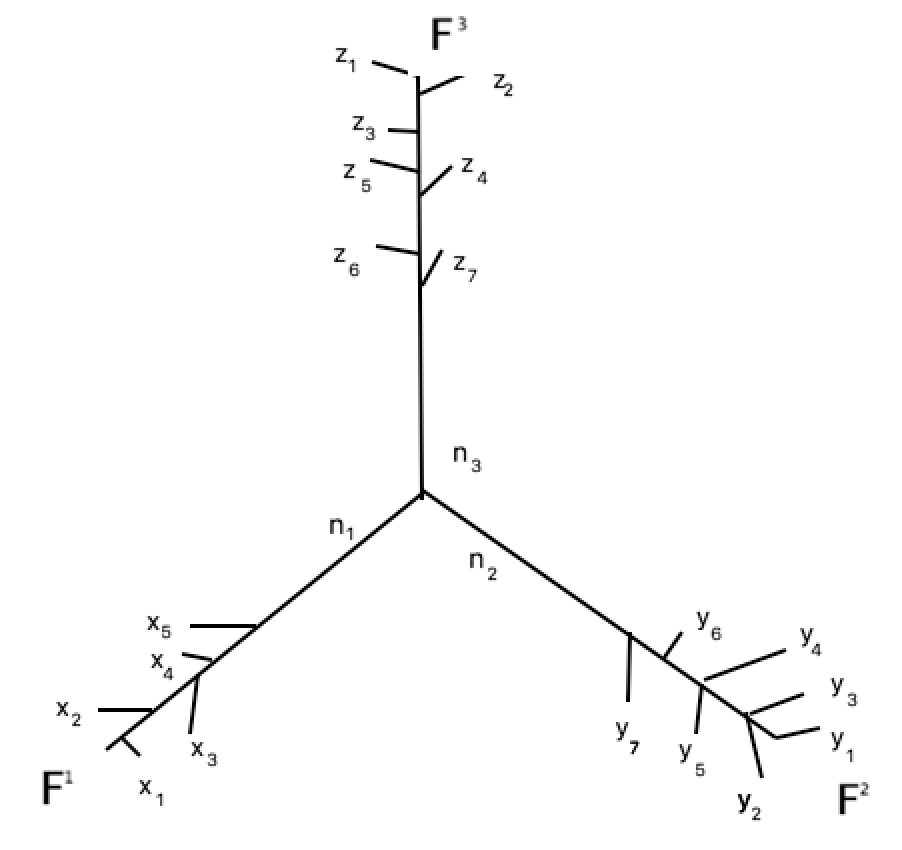}
\caption{The figure gives the abstract picture of the above mentioned decomposition of the vertex. Using such a decomposition, any $n$-valent vertex can be reduced to a $3$-valent one. Indeed,  the edges of the vertex are distributed among three sets $S_1, \,  S_2$ and $S_3$, whose total spin labels are respectively $\sum_i x_i \, , \sum_j y_j$ and $\sum_k z_k$. Thus, each of the sets is recast into a single edge denoted with $n_1$ for, e.g., the $S_1$ set. These three total edges now converge at a $3$-vertex.}
\label{triv}
\end{figure}

\noindent Before we introduce our COs, it is useful to first review how one can associate an operator, in LQG, to angles in space. Consider a sphere around a node $n$, with several edges emanating from it,  on the spin-network graph.  Let us choose three regions on the surface of a sphere and accordingly divide the edges in three sets $S_e$ with $e = \{ 1,2,3 \}$ as in Fig. (\ref{ang}). Given this decomposition of the links, it is always possible to regard a generic node $n$ as a trivalent node \cite{major1}. Here $S_1, S_2$ and $S_3$ refer to the set of edges that meet at $n$, labeled respectively by $1, 2$ and $3$.  Suppose all the edges are outgoing and associate a flux operator $\widehat{F}_i^{e}$  that identifies the direction of each of these sets. In other words, they are the fluxes through the surfaces dual to the (set of) links $S_e$ --- see Fig. (\ref{triv}).  In order to have null angular momentum at the node one has to impose a closure condition $\widehat{F}^1 + \widehat{F}^2 + \widehat{F}^3 = 0$. Then, as first recognized in \cite{major1}, one can define the cosine operator of the angle $\theta$ between $S_1$ and $S_2$ as
\begin{equation}
\widehat{\cos\theta} := \frac{\widehat{F}^1_i \widehat{F}^2_i}{\sqrt{\widehat{F}^1_l \widehat{F}^1_l}\sqrt{\widehat{F}^2_k \widehat{F}^2_k}}   \, .
\end{equation}
Analogously, one can of course define the cosine of the angle between $S_2$ and $S_3$, and for that between $S_1$ and $S_3$.

Its spectrum can be obtained by acting on the spin-network state associated to the graph (\ref{ang}), and by using the closure condition, and is given by
\begin{equation}
\label{cos}
\widehat{\cos\theta} \left | \Psi \right > = \frac{j_3(j_3+1) -j_1(j_1+1) -j_2(j_2+1)}{\sqrt{j_1(j_1+1)}\sqrt{j_2(j_2+1)}} \left | \Psi \right >
\end{equation}
up to a numerical prefactor. Here $j_3$ is the total spin number labeling the group of edges $S_3$, $j_1$ is the total spin of $S_1$ and, finally,  $j_2$ labels $S_2$. As already shown in \cite{major1}, on taking the naive classical limit of this cosine operator, we can regain the cosine of the angle between the two surfaces.

Analogously, we introduce an operator for the sine of the angle as
\begin{equation}
\label{sin}
\widehat{\sin\theta} := \frac{n_i \epsilon_{ijk} \widehat{F}^1_j \widehat{F}^2_k}{\sqrt{\widehat{F}^1_l \widehat{F}^1_l}\sqrt{\widehat{F}^2_k \widehat{F}^2_k}}   \, ,
\end{equation}
where $n_i$ is the normal versor along the internal directions $\{j,k\}$. This operator is defined in a more natural way through the wedge product of two of the fluxes.

The reader should notice that both Eq. \eqref{cos} and Eq. \eqref{sin} make no reference to the space-time manifold but are rather defined only in terms of quantities on the abstract spin-network graph, namely its edges and nodes. We will then try to realize this background independence also for the case of points or coordinates.

\section{Quantized directions imply quantized coordinates}

\noindent Using the sphere described above for the angle operator, around a node, we can separate the surface of the sphere into three regions $S_e$ with $e = \{ 1,2,3 \}$, which like before, collect the edges through each of these regions and assign a flux operator $\widehat{F}_i^{e}$  that labels an outgoing direction for each of these regions. Then, using the outer product of two fluxes, we can define coordinate operators, which need not be orthogonal even in the classical limit. 

The COs are introduced as
\begin{equation}
\label{x}
\widehat{X} := r  \frac{n^i \epsilon_{ijk} \widehat{F}^2_j \widehat{F}^3_k}{\sqrt{\widehat{F}^3_l \widehat{F}^3_l}\sqrt{\widehat{F}^2_k \widehat{F}^2_k}}   \, ,
\end{equation}

\begin{equation}
\label{y}
\widehat{Y} := r \frac{n^i \epsilon_{ijk} \widehat{F}^3_j \widehat{F}^1_k}{\sqrt{\widehat{F}^3_l \widehat{F}^3_l}\sqrt{\widehat{F}^1_k \widehat{F}^1_k}}   \, ,
\end{equation}

\begin{equation}
\label{z}
\widehat{Z} := r \frac{n^i \epsilon_{ijk} \widehat{F}^1_j \widehat{F}^2_k}{\sqrt{\widehat{F}^1_l \widehat{F}^1_l}\sqrt{\widehat{F}^2_k \widehat{F}^2_k}}   \, ,
\end{equation}

where $r$ is a constant with dimensions of a length. Let us first stress that $\widehat{X}$, $\widehat{Y}$, and $\widehat{Z}$ are not usual space-time manifold coordinates, but rather our proposal for a `` notion of coordinate" on the abstract spin-network. Taken one node as reference point, we used the directions of (three of) its links in order to define a 3d basis suitable for introducing objects that resemble usual coordinates.  Thus, the elements of this basis should provide locally the position with respect to a given specific node. Indeed, our generalizations for space directions are identified in terms of the angular momenta of the three groups of edges converging into the same node, which is picked as an origin of the ``coordinate frame" we build. In particular, they are given in terms of the cross product between orthogonal flux operators identifying the three directions of space. 

Let us first note that these COs defined in this way, are naturally regularized in a well-defined sense. Consider each of the circular regions $S_e$ to have a radius $\epsilon$. When we take the limit $\epsilon \rightarrow 0$, both the numerator and the denominator blows up but the CO remains well behaved. To make this more precise, we define the integrated fluxes with smearing functions as done in \cite{major1}
\begin{eqnarray}
[F^e]_f = \int_{S_e} \text{d}^2S \, f^i_\epsilon \, n_a E^a_i\,,
\end{eqnarray}
where $(e=1,2,3)$ stands for the three surfaces\footnote{This method would work even when smearing with different test functions across the three different surfaces.}. In the limit $\epsilon\rightarrow 0$, we have the test function replaced by a delta distribution. Obviously, one can immediately notice that the COs have been defined such that the dependence on the test function, as well as the area of the surfaces, drops out of the expressions \eqref{x}-\eqref{z}. Thus our expressions have already been regularized, in the sense that it is free from the dependence on all of the fiducial structures introduced.

It is important to note that the domains of these operators have to be defined in a suitable way. Since each of these operators have two of the area operators appearing in the denominator, it implies that there has to be at least one edge piercing each of the surfaces on the sphere. In other words, there appears the area operator in the denominator of the these operators. Since the area operator has an eigenvalue for a surface only when an edge of the spin-network graphs intersects it, this would make the CO ill-defined if this would not be the case. Thus, the requirement for the operators to be well-defined should be that the sphere encloses one and one node alone and that each of the surfaces on the sphere must have some edges coming out of them.  

Now that we have discussed a few subtle aspects to take into account, let us underline that this definition we have introduced fits a set of necessary and reasonable requirements. Indeed, given our definitions for quantum coordinates, if one wishes to locate a point on a spin network, this can be done thanks to the above introduced operator, written more compactly as
$$\widehat{R}^{e} = \frac{r \epsilon^{ee'e''} (\widehat{F}^{e'}\wedge \widehat{F}^{e''})}{\sqrt{(\widehat{F}^{e'})^2(\widehat{F}^{e''})^2}},$$
$r$ being the distance of such a point on the classical smooth manifold (in which the spin-network is embedded) from the node. Naturally, one could question what might be an appropriate choice for the value of $r$ appearing in our expressions. Given the above discussion, it should be clear that it is an arbitrary parameter with the dimension of length, whose value depends on the point we refer to. Of course, one might be worried that, being $r$ arbitrarily large, we are introducing an unphysical noncommutativity on large scales then. From this perspective, a natural choice would be taking $r \equiv \lp = \sqrt{\hbar G/c^3}$ --- eventually dependent on the Barbero-Immirzi parameter $\gamma$ as well, if one considers also the details of the lattice regularization adopted. However, it is worth noting that, as we discuss below, the classical limit is recovered in the large spin limit rather than naively sanding $\lp \, \rightarrow \, 0$. Another approach would be finding a way to link $r$ to the length operator described within LQG  \cite{len2,len3}. We will explore such a possibility in Appendix A. In this regards, let us notice that, of course, the construction we display does not represent the only possible definition of operators for coordinates. Instead of starting from Cartesian coordinates, one might use for instance Gaussian normal coordinates \cite{radgaug1,radgaug2,radgaug3}, and try to find a suitable quantization procedure. However, the definition we introduced has the advantage of being closely related to the LQG angle operator.

Our next task is to compute the spectra and, finally, we want to look at the algebra. To this end, we need to act with these COs over spin-network states $\left| \Psi \right > :=  \left| j , m \right >$ of the geometry. Adopting the usual notation, the principal quantum number $j$ labels the irreducible representations of the $SU(2)$ internal gauge group, while $m$ denotes its projection along one of the three available spin directions. Since we desire to show how the semi-classical limit of these COs can be obtained, the best option is to use the so-called coherent-picture of operators recently introduced in Ref.~\cite{CSalesci}. This provides a representation of operators in the basis of semi-classical state vectors. Indeed coherent states are semi-classical spin-networks in the sense that they are peaked on a given classical geometry. Specifically, in spin-foam models it has been shown that these states exponentially dominate the partition function that sum over geometries \cite{CSetera,CSbianchi}, and can also be picked on space-time backgrounds of cosmological interest \cite{Mamape}. Another way of saying that coherent states are semi-classical is that they minimize the uncertainty of phase-space operators. We will briefly comment on this below. Notice that this can be rigorously done since coherent states provide an (over-complete) basis for the kinematical Hilbert space (see e.g. \cite{CSetera}) we are interested in. Let us explicitly specify that our Hilbert space is constructed from the tensor product of three Hilbert spaces (one for each flux $\widehat{F}^e$ defined over the surface $S^e$), i.e. $\mathcal{H}_{tot} := \bigotimes^{3}_{e=1}\mathcal{H}_{e}$ where it is useful to remind that  $\sum^3_{e=1}\widehat{F}^e =0$. Consequently, our space is given by $\mathcal{H}_{tot} \simeq SU(2)\times  SU(2)\times SU(2)$. Let us stress that we are free to choose different quantum numbers $m$ for each of these three Hilbert spaces. Indeed, we will make use of that in order to simplify the computation of the spectrum of our CO later in this section. The starting point is to recognize that coherent states furnish an (over) complete basis of the Hilbert space, { i.e.}
\begin{equation}
\mathbb{I} = \int_\Gamma d \mu (g,\overrightarrow{p}) \left | g, \overrightarrow{p} \right > \left < g,\overrightarrow{p} \right | \, .
\end{equation}
Here $(g,\overrightarrow{p}) \in \Gamma$ identifies a point of the phase-space, $g$ denoting a group element of $SU(2)$ such that $\left<g | j,m \right> = \sqrt{2j+1}\,\overline{\mathcal{D}^j}(g)$, and $\overrightarrow{p}$ standing for the quantum number of momenta. The explicit expression for the Haar measure $d \mu (g,\overrightarrow{p})$ in the coherent-state expansion is given in \cite{CSetera}. We do not report it here since it will not play any role in our analysis. Using this representation of the identity matrix, any operator can be constructed in the following way
\begin{equation}
\widehat{O}_f = \int d \mu \, f(g,\overrightarrow{p}) \left | g, \overrightarrow{p} \right > \left < g,\overrightarrow{p} \right |\,,
\end{equation}
with a proper choice of the functions $f(g,\overrightarrow{p})$. This gives what is called the coherent-state representation of an operator. The $SU(2)$ gauge invariance of coherent state operators has been discussed in some details in Ref.~\cite{CSalesci}. In this regard, it is worth noticing that coherent states are not invariant under $SU(2)$ transformations and, as a result, one needs to make a suitable choice of the function $f(g,\overrightarrow{p})$ in order to obtain a gauge invariant combination for the desired operator $\widehat{O}_f $. For our purposes here, it is of particular relevance the fact that one can introduce a coherent-state picture for the flux operator, which is invariant under left multiplications by $SU(2)$ elements, by identifying $f \equiv \overrightarrow{p}$ \cite{CSalesci}. Indeed, this ensures the gauge invariance of the (coherent-state representation of the) operators \eqref{x}, \eqref{y}, and \eqref{z}. 

The CO depends on flux operators. Thus, in order to compute the action of COs on coherent spin-network states, we only need to know the action of the flux operators. In the coherent-state picture, fluxes can be represented as
\begin{equation}
\widehat{F}^e_i = -i \int d\mu \, p_i \left | g^e, \overrightarrow{p} \right > \left < g^e,\overrightarrow{p} \right | \, ,
\end{equation}
and their (left) action on spin-network states is \cite{CSalesci}
\begin{equation}
\label{cflux}
\widehat{F}^e_i  \left | j^e, m \right> = \frac{i}{2} F_t(j^e) \sigma^{(j_e)} _i  \left | j^e, m \right> \, ,
\end{equation}
where --- see {e.g.} Ref.~\cite{CSalesci} --- the $F_t(j^e)$ coefficient reads
\begin{equation}
\begin{split}
F_t(j_e) = \frac{1}{2t(2j_e+1)j_e(j_e +1)} \Big[ j_e(t(2j_e+1)^2 +2) \\ - \exp \left(-\frac{(2j_e+1)^2t}{4}\right) \sum_s (1+2s^2t)\exp(s^2t) \Big] \, .
\end{split}
\end{equation}
Here $t$ is a parameter that controls the classicality of the coherent states, often called the Gaussian time. Small values of $t$ correspond to states that are sharply peaked on a prescribed geometry of space. For simplicity, let us neglect the normalization in Eq.~\eqref{x}-\eqref{z}. Taking into account Eq.~\eqref{cflux}, for the cross-product operator $\epsilon_{ijk} \widehat{F}^e_j \widehat{F}^{e'}_k$ we can easily find
\begin{eqnarray*}
&&\epsilon_{ijk} \widehat{F}^e_j \widehat{F}^{e'}_k  \left | j^e, m_j \right> \left | j^{e'}, m_k \right> \\ &&= -\frac{\epsilon_{ijk}}{4} F_t(j^e) \sigma^{(j_e)} _j  \left | j^e, m_j \right> F_t(j^{e'}) \sigma^{(j_{e'})} _k  \left | j^{e'}, m_k \right> \, .
\end{eqnarray*}
Retaining the normalization factor $\sqrt{\widehat{F}^e  \widehat{F}^e} \sqrt{\widehat{F}^{e'}  \widehat{F}^{e'}}$, we cannot obtain an analytic expression for the action of the coordinate operators on coherent states, but we can make a numerical integration over the tensor product of the three phase-space corresponding to the three links $S_1,S_2$ and $S_3$. Starting from the above formula, we can compute the algebra closed by the COs. Again, omitting the normalization part of the operators, we calculate the action of the commutation relation
\begin{equation}
 \epsilon_{ijk} \epsilon_{lmn} [ \widehat{F}^e_j \widehat{F}^{e'}_k,  \widehat{F}^{e'}_m \widehat{F}^{e''}_n] \, ,
\end{equation}
over spin-networks associated to trivalent nodes with edges colored with spins $j^e, j^{e'}$ and $j^{e''}$, i.e.
\begin{eqnarray*}
&&\epsilon_{ijk} \epsilon_{lmn}  \widehat{F}^e_j \widehat{F}^{e'}_k \widehat{F}^{e'}_m \widehat{F}^{e''}_n  \left | j^e, m \right> \left | j^{e'}, m \right>  \left | j^{e''}, m \right>  - (\leftrightarrow)\\ &&= \frac{\epsilon_{ijk}\epsilon_{lmn} }{16} F_t(j^e) \sigma^{(j_e)} _j  F_t(j^{e'}) \sigma^{(j_{e'})} _k \\&&\times F_t(j^{e'}) \sigma^{(j_e')} _m  F_t(j^{e''}) \sigma^{(j_{e''})} _n \left| \psi \right>  - (\leftrightarrow) \, ,
\end{eqnarray*}
where, for brevity, we rename
$$ \left| \psi \right> \equiv \left | j^e, m \right> \left | j^{e'}, m \right>  \left | j^{e''}, m \right>\,.$$
Here, the symbol $(\leftrightarrow)$ stands for the second term of the commutator where fluxes are exchanged, namely  the operator $ \epsilon_{ijk} \epsilon_{lmn}  \widehat{F}^{e'}_m \widehat{F}^{e''}_n \widehat{F}^e_j \widehat{F}^{e'}_k$. Then, taking into account that $[\sigma^{(j_e)} _i, \sigma^{(j_{e'})} _j] = 2i \epsilon_{ijk}\sigma^{(j_e)} _k \delta_{j_e \, j_{e'}}$, we find for the commutator
\begin{equation}
\label{calc}
\frac{\epsilon_{lnj}}{8} F_t(j^e)F^2_t(j^{e'})  F_t(j^{e''})\sigma^{(j_{e'})} _i  \sigma^{(j_e)} _j \sigma^{(j_{e''})} _n  \left | \psi \right>  \, .
\end{equation}
Reminding the definition of coordinates \eqref{x}, \eqref{y} , \eqref{z} and using the above calculation \eqref{calc}, we can write down the commutators between coordinate operators. We find the following algebra
\begin{equation}
\begin{split}
&[\widehat{X},\widehat{Y}] = i \widehat{Z}\frac{\widehat{F}^3}{(\widehat{F}^3)^2} \, , \quad [\widehat{Z},\widehat{X}] = i \widehat{Y}\frac{\widehat{F}^2}{(\widehat{F}^2)^2} \, , \\ &[\widehat{Y},\widehat{Z}] = i \widehat{X}\frac{\widehat{F}^1}{(\widehat{F}^1)^2}  \, ,
\end{split}
\end{equation}
having omitted the internal indexes. Here we have also used the fact that flux operators belonging to different edge sets commute, namely
\begin{equation}
[\widehat{F}^e_i, \widehat{F}^{e'}_j ] = 0 \, , \quad e \neq e' \, ,
\end{equation}
 and that we are considering orthogonal edge directions 
 \begin{equation}
 \widehat{F}^e_k \widehat{F}^{e'}_k = 0  \, , \qquad e \neq e' \, .
\end{equation}
We obtained a noncommutative algebra for our COs, in which the associative property is still preserved. Indeed, we can write down the Jacobi identity, namely 
\begin{equation}\label{jacobi}
\begin{split}
[[\widehat{X},\widehat{Y}],\widehat{Z}] + [[\widehat{Z},\widehat{X}],\widehat{Y}] + [[\widehat{Y},\widehat{Z}],\widehat{X}] =\\  [\widehat{Z}\frac{\widehat{F}^3}{(\widehat{F}^3)^2}, \widehat{Z}] + [\widehat{Y}\frac{\widehat{F}^2}{(\widehat{F}^2)^2} ,\widehat{Y}] + [\widehat{X}\frac{\widehat{F}^1}{(\widehat{F}^1)^2}, \widehat{X}] \equiv 0 \, ,
\end{split}
\end{equation}
where we have used the fact that $\widehat{Z}$ commutes with $\widehat{F}^3$, since it depends only on the other two fluxes. An analogous observation applies to the other two commutators in the above expression. The first comment that is worth making at this point is that COs do not commute, as a consequence of the LQG quantization. Bearing in mind the form of coordinate operators that are expressed in terms of fluxes --- namely Eqs.~\eqref{x}, \eqref{y}  and \eqref{z} ---  it is possible to understand this noncommutativity as a direct consequence of having an internal $SU(2)$ symmetry. The noncommutativity  can be seen as arising from the quantization of the $SU(2)$ Poisson brackets. Furthermore, it is worth commenting the fact that the algebra of coordinates we have derived closely resembles the commutation relations for the fuzzy sphere \cite{madore}. In fact, the above commutators can be succinctly rewritten as
\begin{equation}
\label{comm}
[\widehat{X}^e, \widehat{X}^{e'} ] = i \epsilon^{ee'e''} \widehat{X}^{e''} \frac{\widehat{F}^{e''}}{(\widehat{F}^{e''})^2} \, ,
\end{equation}
where the indexes refer to the three edge directions identified by $S_1$, $S_2$ and $S_3$. The main difference with respect to the standard fuzzy-sphere commutators resides in the appearance of more complicated structure functions (rather than structure constants) in our case \eqref{comm}.  The interest for the fuzzy sphere comes from the fact that it is the noncommutative algebra of space coordinates that arises in  $3D$ quantum gravity \cite{freidelivine}. However, we do not obtain exactly the algebra of the fuzzy sphere due to the fact that on the right-hand side of the commutator there is still an explicit dependence on the flux. Nonetheless, our result provides a first constructive realization of the ideas of noncommutative geometry from LQG.\\

Finally we show that the classical commutative limit can be recovered in the large spin approximation. To this end let us compute the action of the commutator \eqref{comm} on a generic spin-network state associated to our 3-vertex, which we formally write as $\left | \Psi \right > = \left | j_{e} , m_{e} \right > \left | j_{e'} , m_{e'} \right >\left| j_{e''} , m_{e''} \right >$. For simplicity, let us make the case with $e = 1$, $e' = 2$, and $e'' = 3$. Thus, we are taking a spin-network states given by the tensor product of three holonomies related to the three different edges of our vertex. Let us expand two holonomies in the internal $z$-direction and one on the internal $x$-direction, i.e. $ \left | \Psi \right > = \left | j_{1} , m^z_{1} \right > \left | j_{2} , m^x_{2} \right >\left| j_{3} , m^z_{3} \right >$.  Then, the action of the commutator $[\widehat{X}^1, \widehat{X}^{2} ]$ reads
\begin{equation}
\label{class}
[\widehat{X}^1, \widehat{X}^{2} ] \left | \Psi \right > =  \frac{i\delta_{lx} m^z_1m^z_3 m^x_2}{\sqrt{j_1(j_1+1)}\sqrt{j_2(j_2+1)}j_3(j_3+1)} \left | \Psi \right > \, ,
\end{equation}
having neglected numerical overall factors. From the above equation the reader can easily recognize that the classical limit coincided with the large spin limit with $j_3 \rightarrow \infty$, which restores the commutativity of coordinates. In Eq.~\eqref{class}
$$ \frac{m_1m_3 m_2}{\sqrt{j_1(j_1+1)}\sqrt{j_2(j_2+1)}\sqrt{j_3(j_3+1)}}\sim\mathcal{O}(1)$$ 
and, then, we have a factor  $1/\sqrt{j_3(j_3+1)}$ that involves the spin on the internal edge $S_3$ shared by both $\widehat{X}$ and $\widehat{Y}$. The classical limit corresponds to the requirement of having large spins on the internal direction $S_3$ and, as desired, for large values of $j_3$ the right hand side of Eq. \eqref{class} collapses to zero. The fact that the (semi-) classical limit can be obtained by taking the large spin limit lies at the very root of the role of coherent states and their role in bridging classical and quantum regimes \cite{CSetera,CSbianchi,Mamape,CSalesci}. A different coarse-graining method for LQG states has been recently proposed in Ref.~\cite{norbert1}, where, instead of increasing the spin number, one increases the number of vertices while keeping fixed the total volume in order to reach a semi-classical continuum limit. Indeed, it might be insightful analyzing the behavior of our operators under this limit but it will be explored elsewhere.

\section{Conclusion and outlook}

\noindent
Moving from a microscopic theory of background independent quantization of gravity, LQG, we found a path to derive at intermediate (mesoscopic) scales a noncommutative geometric structure of the semiclassical geometry that are reminiscent of the fuzzy sphere.

Specifically, we presented a proposal for coordinate operators in LQG. Our construction relies on some properties of operators for angles that were already established in the literature. The definition of the COs analyzed here has been instantiated in a background-independent fashion, and the action of the operators has been automatically specified on the kinematical Hilbert space of LQG. Thus this set of operators we discussed in this paper has been tailored to account for the quantum texture of the geometry at the Planck scale. The grouping of edges in a finite amount of sets \cite{major1}, which is preliminary to the definition of these operators in our work, together with the computation of the action of these operators on coherent states, played a crucial role in our working strategy. Indeed these steps enabled us to develop a coarse-graining and semiclassical procedure that unveiled the noncommutativity of the spatial coordinates at mesoscopic scales. Finally, extracting the large $j$ limit out of the action of the operators on the coherent states, it has been possible to recover the coordinates' commutativity of the space-time manifold on macroscopic scales.

There are several aspects that still need to be explored further in order to strengthen the picture we propose here. In particular it is meaningful to ask whether the algebra of coordinates we recovered hitherto may acquire a dependence on the chosen topology for the manifold our consideration starts from. Furthermore, we did not address yet in this work the reconstruction of the algebra of symmetries dual to the noncommutative version of space-time we recovered. Because of the Hamiltonian analysis involved, we expect that at least it should be possible to look at the subgroups of translations and spatial rotations of the Poincar\'e group, and then comment on their (eventually expected) deformation. We emphasize that noncommutative coordinates that are encountered in some examples of noncommutative geometry do not retain the interpretation of "COs". Nonetheless the use of noncommutative coordinates, intended as auxiliary labels, is dual to the deformation of the space-time algebra of symmetries. The latter is usually cast in the framework of quantum groups, and has a clearer physical meaning at the level of some testable predictions of the theory (e.g., modified energy-momentum dispersion relations for particles). Finally, we are interested in understanding whether (and how) properties such as homogeneity or isotropy would affect the algebra of our operators. For instance,  taking the same quantum coordinates at each nodes would tell us that the graph is homogeneous (or, in other words, a regular lattice) with the same quantum numbers labeling all the edges.

We emphasize that our result, although preliminary,  sheds some light on the role of noncommutative geometry in LQG. In particular, the findings of this paper gives further support to the idea that noncommutativity can be then understood as an effective arena derived from LQG at mesoscopic scales, while space-time commutativity can be still recovered at large (macroscopic) scales.

Once an analysis of the dual symmetries will be properly developed,  it will be tempting to apply to the scheme discussed here recent phenomenological considerations on the infrared regime for quantum gravity and the Newton constant. Very severe constraints have been derived on space-time noncommutativity looking at underground experiments, searching for violations of the Pauli exclusion principle \cite{Addazi:2017bbg}. For some notable cases, including $\kappa$-Minkowski and $\theta$-Minkowski, taking into account natural values of the noncommutativity scale involved, space-time noncommutativity can be already ruled out. In general, tight constraints can be recovered up to the third order in the ratio between the particles' energies involved in the nuclear processes that are measured, and the Planck energy scale (defined in the infrared limit). Consequently, it turns to be crucial deepening the dual structure of the space-time symmetries, in order to unveil whether violations of the Pauli exclusion principle are present in these emergent non-commutative frameworks. This might provide a superselection principle to be applied to the possible variants through which space-time noncommutativity can be accounted for in LQG. 

\section*{ACKNOWLEDGEMENTS} 
\noindent 
A. M. and M. R. are grateful to Giorgio Immirzi for reading a preliminary version of this work, and acknowledge his support and encouragement. S. B. and A. M acknowledge comments and suggestions from Jerzy Lewandowski. M. R. thanks Eugenio Bianchi for his useful comments and suggestions. The authors also thank Norbert Bodendorfer for reading a previous version of the manuscript and suggesting us some improvements.  A. M. wishes to acknowledge support by the Shanghai Municipality, through the grant No. KBH1512299, and by Fudan University, through the grant No. JJH1512105. The contribution of M. R. is based upon work from COST Action MP1405 QSPACE, supported by COST (European Cooperation in Science and Technology). 

\appendix 

\section{Quantized lengths imply quantized coordinates}
\noindent
In this Appendix we show that it is possible to reproduce a similar result than the one exposed in the previous sections, about the emergence of space-time noncommutativity at mesoscales. This is achieved by modifying slightly the construction for the COs we have given in the main text. The result we are going to provide below clearly shows that an ambiguity might arise in the constructive method we adopted, while determining the particular kind of emergent space-time noncommutativity. Nevertheless, the result we derive in this appendix, making use of slightly different technical details with respect to the main text, confirms the emergence of space-time noncommutativity at mesoscales from LQG, and then exhibits a certain solidity of our observation. 

We start by writing explicitly the operators along the unit vectors that can be decomposed on the three coordinate axes: \begin{equation}
\label{ix}
\widehat{i_X} :=  \frac{ n^i \epsilon_{ijk} \widehat{F}^2_j \widehat{F}^3_k}{\sqrt{\widehat{F}^3_l \widehat{F}^3_l}\sqrt{\widehat{F}^2_k \widehat{F}^2_k}}   \, ,
\end{equation}

\begin{equation}
\label{iy}
\widehat{i_Y} :=  \frac{ n^i \epsilon_{ijk} \widehat{F}^3_j \widehat{F}^1_k}{\sqrt{\widehat{F}^3_l \widehat{F}^3_l}\sqrt{\widehat{F}^1_k \widehat{F}^1_k}}   \, ,
\end{equation}

\begin{equation}
\label{iz}
\widehat{i_Z} :=  \frac{ n^i \epsilon_{ijk} \widehat{F}^1_j \widehat{F}^2_k}{\sqrt{\widehat{F}^1_l \widehat{F}^1_l}\sqrt{\widehat{F}^2_k \widehat{F}^2_k}}   \, .
\end{equation}
In this way, space directions are identified in terms of (the cross product of) the angular momenta of the three sets of edges emanating from the same node. Each of these cartesian coordinates (i.e. $\widehat{i_X}, \widehat{i_Y}$ and $\widehat{i_Z}$ ) should be thought as a unit vector along each of these (local) coordinate axes. The choice of a cube ensures that this local coordinate system is orthogonal in the classical limit. The requirement of having orthogonal links -- $ \widehat{F}^e_k \widehat{F}^{e'}_k = 0$ if $e \neq e'$ -- is then automatically satisfied. 

\begin{figure}[h!]
\centering
\includegraphics[width=3in]{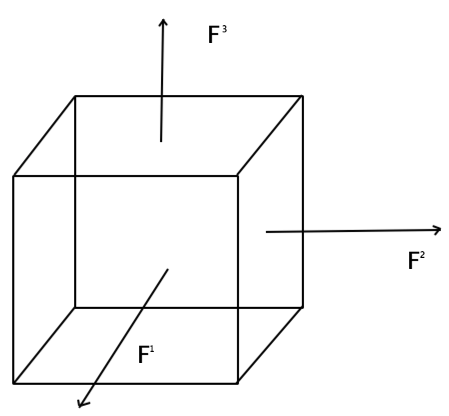}
\caption{The figure shows our decomposition of the fluxes associated to the edges converging at a given node of the spin-network. Just as we did already for quantum directions, i.e. $\widehat{i}_X, \widehat{i}_Y$ and $\widehat{i}_Z$, we can group edges in three different sets. Then, we associate to each of these three sets a total flux $\widehat{F}^e$ with $e=1,2,3$. However, we are here choosing a different topology for our edge decomposition. Taking a cube ensures that the coordinate basis we define is orthogonal and we do not need to impose any orthogonality conditions on fluxes. }
\label{ang1}
\end{figure}

Our aim is to define COs which have dimensions of length in order to avoid to add by hand the $r$ parameter. To this aim, we can use the expression for the length operator from LQG. From Ref.~\cite{len3}, we have that
\begin{eqnarray}\label{Lx}
  L_x = \frac{\text{Ar}(S_2)\text{Ar}(S_3)|\sin\theta^{23}|}{V(\text{cube})}\,.
\end{eqnarray}
$S_2, S_3$ are the two faces of the cube which are orthogonal to the face $S_1$, the latter collecting all the edges corresponding to $\widehat{F}^{1}_k$. The volume in the denominator corresponds to the that of the cube. When we want to turn the above expression into an operator on the LQG Hilbert space, we can now easily do so since we have well-defined operators corresponding to the area, the volume and the sine operator. Obviously, analogous expressions hold for the other directions as well.

The relation in \eqref{Lx} can be recast in terms of the flux operators as
\begin{eqnarray}\label{lx1}
  \widehat{L_X} = \widehat{V^{-1}} n^l \epsilon_{lmn}\widehat{ F}^2_m \widehat{ F}^3_n\,.
\end{eqnarray}
Similarly, one can write down the length operators along the $\widehat{i_Y}$ and $\widehat{i_Z}$ directions. These would involve the other fluxes for each of the other length operators. Now we can define the COs as
\begin{eqnarray}\label{xprime}
 \widehat{X} &:=& \widehat{L_X} \widehat{i_X} = \widehat{V^{-1}} \frac{(n^l \epsilon_{lmn} \widehat{F^m_2} \widehat{F^n_3})^2}{\sqrt{\widehat{F^2_j}\widehat{F^2_j}}\sqrt{ \widehat{F^3_k}\widehat{F^3_k}}}\,\\
\widehat{Y} &:=& \widehat{L_Y} \widehat{i_Y} = \widehat{V^{-1}} \frac{(n^l \epsilon_{lmn}\widehat{ F}^m_3\widehat{ F}^n_1)^2}{\sqrt{\widehat{F^3_j}\widehat{F^3_j}}\sqrt{ \widehat{F^1_k}\widehat{F^1_k}}}\,,\label{yprime} \\
\widehat{Z} &:=& \widehat{L_Z} \widehat{i_Z} = \widehat{V^{-1}} \frac{(n^l \epsilon_{lmn}\widehat{ F}^m_1\widehat{ F}^n_2)^2}{\sqrt{\widehat{F^1_j}\widehat{F^1_j}}\sqrt{ \widehat{F^2_k}\widehat{F^2_k}}}\,.\label{zprime}
\end{eqnarray}

The regularization is more subtle in this case, since we have not only the area operator appearing in the denominator, but also the volume operator. For the inverse volume operator, we can follow the exact steps as in \cite{len3}, in order to regularize the operator. We can define the inverse volume operator as
\begin{eqnarray}
  \widehat{V^{-1}} := \lim_{\epsilon \rightarrow  0} \left(\hat{V}^2 + \epsilon^2 \lp^6 \right)^{-1} \hat{V}\,.
\end{eqnarray}
We can assume that there is only one node inside the cube, on which the inverse volume operator acts. This ensures that the inverse volume operator remains well defined in this case and gives a non-zero result when acting on the spin-network states. One way to extend the domain of these operators would be to apply an analogous procedure for the `inverse' area operators. In LQG, the area operator has a discrete spectrum with a non-zero minimum eigenvalue. However, since the denominator goes to zero if none of the edges intersect the surface corresponding to the area operator appearing in the definition, we might say that we are including zero as a discrete eigenvalue of the area operator and then regularizing this inverse operator \`a la Tikhonov, as for the volume operator. This has already been discussed in \cite{len2}, demonstrating that this procedure is quite general and allows us to define a similar procedure for the area operator as
\begin{eqnarray}
  \widehat{A^{-1}} := \lim_{\epsilon \rightarrow  0} \left(\hat{A}^2 + \epsilon^2 \lp^4 \right)^{-1} \hat{A}\,,
\end{eqnarray}
where $\hat{A}$ stands for any of the three area operators. Unlike the case of the inverse volume operator, which necessarily takes a non-zero eigenvalue due to the requirement of a node appearing in the fiducial cube, in this case the inverse area operator can be zero depending on whether there are edges piercing the relevant surface. This extends our definition of the CO since now, if there are no edges coming out of $S^1$, we shall get some of our COs to be zero, whereas $\hat{X}$ shall be nonzero. In this way, we do not require that there are some edges piercing all of the faces of our fiducial cube.

Following similar steps to those reported in the main text, one can compute the algebra of these operators with a tedious but rather straightforward calculation 
\begin{equation}\label{comm1}
\begin{split}
[\widehat{X},\widehat{Y}] = i\widehat{L}_X \widehat{L}_Y (\widehat{L}_Z)^{-1}\widehat{Z} \frac{\widehat{F}^3}{(\widehat{F}^3)^2}\\ +i \widehat{L}_X\widehat{L}_Z \frac{\widehat{F}^3}{\sqrt{(\widehat{F}^2})^2\sqrt{(\widehat{F}^3})^2} (\widehat{L}_Y)^{-1}\widehat{Y} \\+ i\widehat{L}_Y\widehat{L}_Z\frac{\widehat{F}^3}{\sqrt{(\widehat{F}^1})^2\sqrt{(\widehat{F}^3})^2} (\widehat{L}_X)^{-1}\widehat{X}\\+ i\widehat{V}^{-1}\widehat{L}_Z\widehat{F}^3 (\widehat{L}_X)^{-1}\widehat{X} (\widehat{L}_Y)^{-1}\widehat{Y} \, ,
\end{split}
\end{equation}

and

\begin{equation}\label{comm2}
\begin{split}
[\widehat{Y},\widehat{Z}] = i\widehat{L}_Y\widehat{L}_Z (\widehat{L}_X)^{-1}\widehat{X}\frac{\widehat{F}^1}{(\widehat{F}^1)^2}\\ +i\widehat{L}_Y\widehat{L}_X  \frac{\widehat{F}^1}{\sqrt{(\widehat{F}^1})^2\sqrt{(\widehat{F}^3})^2}(\widehat{L}_Z)^{-1}\widehat{Z} \\+i\widehat{L}_Z\widehat{L}_X  \frac{\widehat{F}^1}{\sqrt{(\widehat{F}^1})^2\sqrt{(\widehat{F}^2})^2} (\widehat{L}_Y)^{-1}\widehat{Y}\\+ i\widehat{V}^{-1}\widehat{L}_X \widehat{F}^1 (\widehat{L}_Z)^{-1}\widehat{Z} (\widehat{L}_Y)^{-1}\widehat{Y} \, ,
\end{split}
\end{equation}

and finally

\begin{equation}\label{comm3}
\begin{split}
[\widehat{Z}, \widehat{X}] =    i\widehat{L}_Z \widehat{L}_X (\widehat{L}_Y)^{-1}\widehat{Y} \frac{\widehat{F}^2}{(\widehat{F}^2)^2}\\ +i \widehat{L}_Z\widehat{L}_Y \frac{\widehat{F}^2}{\sqrt{(\widehat{F}^2})^2\sqrt{(\widehat{F}^1})^2} (\widehat{L}_X)^{-1}\widehat{X} \\+ i\widehat{L}_X\widehat{L}_Y\frac{\widehat{F}^2}{\sqrt{(\widehat{F}^2})^2\sqrt{(\widehat{F}^3})^2} (\widehat{L}_Z)^{-1}\widehat{Z}\\+ i\widehat{V}^{-1}\widehat{L}_Y\widehat{F}^2 (\widehat{L}_X)^{-1}\widehat{X} (\widehat{L}_Z)^{-1}\widehat{Z}   \, .
\end{split}
\end{equation}

Here we used the fact that $\widehat{i}_X = (\widehat{L}_X)^{-1}\widehat{X}$, and similar relations for the remaining two directions. The main subtle issue is represented by ordering ambiguities as usual. It is possible to show that also coordinates defined as in Eqs.~\eqref{xprime}, \eqref{yprime}, and \eqref{zprime} have discrete spectra, meaning that positions on a spin-network configuration can be localized only with a finite resolution given by (the inverse of) the minimum eigenvalue. Even without computing spectra explicitly, we can still say something about the behavior of these operators in the large spin $j$ limit. As aforementioned, this is an important consistency check since, in that limit, we expect to recover standard commutative properties. To this end, let us make a rough estimate of the spin order of each term appearing on the right-hand side of Eqs. \eqref{comm1}, \eqref{comm2}, and \eqref{comm3}. In particular, let us do that for the commutator $[\widehat{X},\widehat{Y}]$. One can immediately realize that the following considerations apply directly also to the other two commutators. Given Eq. \eqref{lx1} and taking into account calculations in the previous section, it is not difficult to see that $\widehat{L}_X \, \sim \sqrt{j}$. Thus, it is worth saying that the length by itself is not a well-behaved operator when $j \, \longrightarrow \, \infty$. Indeed, its semi-classical limit is obtained by sending to zero the lattice parameter, i.e. when $\lp \, \longrightarrow \, 0$ --- see, however, Refs.~\cite{len2,len3}. Then, we have $\widehat{Z} \, \sim \, 1$ and, finally, the structure function $\widehat{F}^3/(\widehat{F}^3)^2 \, \sim \, j^{-1}$. In light of this, we have 
$$[\widehat{X},\widehat{Y}] \, \sim \, 1/\sqrt{j} \, \xrightarrow[j\longrightarrow\infty]{} \, 0\,.$$

Finally, we can show once again how to recover the standard space-time coordinates of the manifold in the semi-classical limit. We have that

\begin{equation}
\begin{split}
X = L_X i_X = V^{-1}n^i_1 \epsilon_{ijk}F^j_2 F^k_3 i_X \\ \simeq  V^{-1}n^i_1 \epsilon_{ijk} \delta^4 (n_1 \cdot i_1 ) \,i_1  \, ,
\end{split}
\end{equation}
where we used above results to approximate fluxes. Now, in terms of fluxes, the volume is given by $V = \sqrt{|  \epsilon^{abc}\epsilon_{ijk}F^i_a F^b_j F^c_k |/(3!)}$ and, with $E^a_i \sim \delta^a_i$, we have
$$V \simeq \delta^3 \sqrt{(\delta^i_i\delta^j_j - \delta^i_j \delta^j_i)/(3!)} \equiv \delta^3\,.$$
Plugging it into the above expression for $X$, we then obtain
\begin{equation}
X \simeq \delta\, i_1 \, ,
\end{equation}
having used the fact that $||n|| \equiv 1$ or, equivalently, $n_1 \equiv i_1$. Moreover, at least upon identifying $r \equiv \delta$, which certainly holds locally, the classical and flat limit gives the same result we obtained in the main text for the COs in Eqs. \eqref{x}, \eqref{y}, \eqref{z}. 

\section{Target manifold and naive classical limit}
\noindent
In this section we wish to show, in the simplest way, how the operators for coordinates acting on spin-network states can be related to usual coordinates on a smooth manifold. We have already shown how it is possible to recover the commutative property by taking the large spin $j$ limit of the expectation value of commutators calculated over the coherent states of LQG. Here we just give a naive derivation of standard positions on a manifold, starting from our nodes' COs. In fact, according to the background-independence philosophy, Eqs. ~\eqref{x}, \eqref{y}, and \eqref{z}, as well as Eqs.~\eqref{xprime}, \eqref{yprime}, and \eqref{zprime}, do not identify positions on a manifold but rather on an abstract spin-network graph. In full LQG, we should not make use of the concept of manifold, which is substituted by abstract spin networks. For this reason, we defined our operators only in terms of nodes, edges, and links. However, it is also well known that, at least in the (semi) classical limit one requires the existence of a background manifold in which to embed the spin-network graphs. 

If we take the classical limit naively and ignore all the ordering issues present in the definition of the operators, we can recover geometrical quantities defined on standard manifolds. Then, assuming that triad operators only act in a small region, we can approximate fluxes in \eqref{flux} as $F^{i} \, \approx \, \delta^2 n^a E^{i}_a$ , being then $E^{i}_a$ constant over a small surface $S \, \sim \, \delta^2$ with normal $n^a$. For the sake of brevity and simplicity, we also restrict to sufficiently small $S$ such that curvature is zero. Thus, we have simply $E^a_i \, = \,  \sqrt{h}e^a_i \, \simeq \, \delta^a_i$, where $e^a_ie^b_j \eta^{ij}= h^{ab}$. Under these approximations, let us consider e.g. our former definition for $\widehat{X}$ \eqref{ix} that becomes
\begin{equation}
\begin{split}
X \, \simeq \, r \frac{\epsilon_{ijk}n^a_2E^j_a n_3^b E^k_b}{\sqrt{n_2^aE^i_a n_2^c E^i_c}\sqrt{n^b_3E^i_b n^d_3 E^i_d}} \\ \simeq \, r \frac{\epsilon_{ijk}n^a_2\delta^j_a n_3^b \delta^k_b}{\sqrt{n_2^a\delta^i_a n_2^c \delta^i_c}\sqrt{n^b_3\delta^i_b n^d_3 \delta^i_d}} \, = \, r\frac{\epsilon_{abc}n^b_2 n_3^c}{\sqrt{n_2^e n_2^e}\sqrt{n^d_3 n^d_3}}\\ = r\frac{n_2 \wedge n_3}{||n_2|| ||n_3||} = r\, (i_2 \wedge i_3) = r \, i_1 \, ,
\end{split}
\end{equation}
being $i_1$, $i_2$, and $i_3$ the orthogonal unit vectors providing the directions of the $X$, $Y$ and $Z$ axes respectively. Of course, similar conclusions apply to the operators $\widehat{Y}$ and $\widehat{Z}$ as defined in \eqref{y} and \eqref{z} respectively.  In this limit, we have found meaningful formulas for usual space-time coordinates on a manifold.


\begin{thebibliography}{99}
\bibitem{Smo17} L.\ Smolin, \tia{What are we missing in our search for quantum gravity?} \arX{1705.09208}.
\end{thebibliography}
\end{document}